# A Novel Method of Subgroup Identification by Combining Virtual Twins with GUIDE (VG) for Development of Precision Medicines

Jia Jia, Qi Tang, Wangang Xie, Richard Rode

**Abstract**

A lack of understanding of human biology creates a hurdle for the development of precision medicines. To overcome this hurdle we need to better understand the potential synergy between a given investigational treatment (vs. placebo or active control) and various demographic or genetic factors, disease history and severity, etc., with the goal of identifying those patients at increased "risk" of exhibiting clinically meaningful treatment benefit. For this reason, we propose the VG method, which combines the idea of an individual treatment effect (ITE) from Virtual Twins (Foster, et al., 2011) with the unbiased variable selection and cutoff value determination algorithm from GUIDE (Loh, et al., 2015). Simulation results show the VG method has less variable selection bias than Virtual Twins and higher statistical power than GUIDE Interaction in the presence of prognostic variables with strong treatment effects. Type I error and predictive performance of Virtual Twins, GUIDE and VG are compared through the use of simulation studies. Results obtained after retrospectively applying VG to data from an Alzheimer's disease clinical trial also are discussed.

# Introduction

The concepts of personalized medicine and precision medicine have "evolved" over time, with precision medicine now viewed as an approach that allows for the treatment of patients while taking their personal signatures, such as genes, environment and lifestyles, into consideration in order to maximize the benefit (efficacy) and/or minimize the risk (safety) they receive from the treatment. In other words, treatment effects are often heterogeneous in a given patient population. Thus, it is necessary to improve our understanding of differential treatment effects observed in patients with different signatures.

Although it is difficult to accurately assess and maximize the treatment effect for every patient, it may be possible to categorize patients into subgroups according to some known and pre-defined signatures, for example, demographics, biomarkers and lab values and then assess the treatment effect for those subgroups. The subgroup in which the "best" treatment effect can be observed also could be identified via complicated statistical methods without predefining the variables to be used in the analysis. In this paper, we focus on the latter approach, also known as retrospective (or ad hoc) subgroup identification. Retrospective subgroup identification is a critical approach used to develop precision medicines.

Many subgroup identification methods have been developed. For example, Negassa et al. (Negassa, et al., 2005) developed RECPAM which attempts to maximize the Cox partial likelihood. Su et al. (Su, et al., 2009) (Su, et al., 2008) developed Interaction trees (IT), which attempts to minimize the p-value for testing the significance of the interaction term between the subgroup indicator and the treatment. Foster et al. (Foster, et al., 2011) developed Virtual twins (VT) which uses random forests to predict the treatment effect for each patient and then applies Classification and Regression Trees (CART) to identify potential subgroups. Lipkovich et al. (Lipkovich, et al., 2011) developed the Subgroup Identification based on Differential Effect Search (SIDES), which targets on the treatment effect difference but may lead to selection bias associated with variables having more possible cut-off values. Dusseldorp and Van Mechelen (Dusseldorp & Mechelen, 2014) developed QUalitative INteraction Trees (QUINT) that attempts to optimize a weighted sum of measures of effect size and subgroup size. Loh et al. (Loh, 2002) developed a method called Generalized, Unbiased, Interaction Detection and Estimation (GUIDE) that is a multi-purpose machine learning algorithm for constructing classification and regression trees. Also, Loh et al (Loh, et al., 2015) later developed a new method called GUIDE-Interaction (Gi), which was based on the original GUIDE but targeted on the treatment effect difference. Also, Gi has been compared to other methods regarding the statistical properties in certain scenarios. The results showed that Gi was a preferred solution when the goal is to find the signature based on treatment difference.

In precision medicine, the treatment effect difference forms the basis for subgroup identification. Most of the existing methods can identify subgroups, and some of the methods can obtain unbiased signature selection. However, only a few of these methods can target directly on the treatment effect difference, especially the individual treatment effect (ITE) difference, which in general is the benefit obtained from receiving treatment compared to the benefit obtained from receiving placebo for a particular patient.

During our review of the existing methods, we found that the variable selection process of Gi is partly driven by how well a variable predicts the response variable. Thus, in a case of a strong prognostic effect, Gi may select prognostic variables more often than predictive variables, since prognostic variables may in fact predict the response variable better than predictive variables. In contrast, VT directly targets on the treatment effect difference and may have better variable selection performance than Gi in the scenario of a strong prognostic effect. However, CART is implemented in the variable selection step of VT and may lead to bias in variable selection (Loh, et al., 2015).

In this paper, we propose VG, a novel method that targets directly on individual treatment effects using an unbiased variable selection procedure by combining two methods, VT and GUIDE. In VG, the CART part in VT is replaced by GUIDE. The performance of VG, VT and Gi will be compared via simulations and a case study will be presented.

**Methods**

The VG method contains two steps: (1) estimate the individual treatment effect (ITE) difference using Virtual Twins; and (2) identify potential subgroup(s) using GUIDE. Benefits inherited from GUIDE include the ability to utilize missing covariate information and simultaneously model multiple endpoints.

Without loss of generality, we illustrate the VG method for the case of binary response variable.

**Step I.**

The first step is to estimate the ITE difference by using Random Forests (Foster, et al., 2011). Let $Y_n$ and $T_n$ represent the original response variable (continuous or binary) and treatment variable (0 = placebo; 1 = investigational treatment), respectively; and let $X_{n,p}$ represents the matrix that contains all the covariates, where n is the sample size, and p is the number of covariates.

Let $T'_n$ represent the flipped (or opposite) treatment variable, where $T'_n = 1_n - T_n$. The purpose of doing this is to estimate the counterfactual response of each patient, in other words, the response of a patient under the treatment that was not received.

Let $Y'_n$ represent the estimated counterfactual response given $T'_n$ and $X_{n,p}$. In the VG method, GUIDE is used to provide nonparametric estimation of $Y'_n$. In order to do that, similar to Foster et al. [reference], we utilize two additional matrices that contain all possible two-way interactions between (a) $T_n$ and $X_{n,p}$ and (b) $T'_n$ and $X_{n,p}$, respectively.

As a consequence, data used for predicting $Y'_n$ have the structure provided below, which contains five components and a total sample size of 2n:

$$Y^*_{2n} = \begin{pmatrix} Y_n \\ Y'_n \end{pmatrix}, \quad T^*_{2n} = \begin{pmatrix} T_n \\ T'_n \end{pmatrix}, \quad X^*_{2n,p} = \begin{pmatrix} X_{n,p} \\ X_{n,p} \end{pmatrix}, \quad XT_{2n,p} = \begin{pmatrix} T_n X_{n,p} \\ T'_n X_{n,p} \end{pmatrix},$$

$$X(1-T)_{2n,p} = \begin{pmatrix} (1_n - T_n)X_{n,p} \\ (1_n - T'_n)X_{n,p} \end{pmatrix}$$

The terms $T_n X_{n,p}$, $T'_n X_{n,p}$, $(1_n - T_n)X_{n,p}$ and $(1_n - T'_n)X_{n,p}$ represent interactions between the treatment indicator, flipped treatment indicator and the covariates. Note that $Y'_n$ is unknown but can be predicted using GUIDE with a weight variable, which assigns 0 weight to $Y_n$ and $1_n$ weight to $Y'_n$.

After $Y'_n$ is predicted, we can calculate the individual treatment effect (ITE) difference. Let's assume the i$^{th}$ patient with individual covariates $X_i$ received treatment ($t_i$= 1) and obtained outcome $y_i$, and the j$^{th}$ patient with individual covariates $X_j$ received placebo ($t_j$= 0) obtained outcome $y_j$. By predicting $Y'_n$, we have now obtained the 'flipped' outcome for patients i and j, $y'_i$ and $y'_j$, given their individual covariates $X_i$ and $X_j$, respectively. Thus, the ITE difference for the i$^{th}$ patient can be calculated as $ITE_i = y_i - y'_i$; while the ITE difference for the j$^{th}$ patient can be calculated as $ITE_j = y'_j - y_j$. By doing this, we obtain the vector of length n containing the ITE differences for all the patients, conditional on their individual covariates.

**Step II.**

The goal of Step II is to identify treatment effect heterogeneity based on the estimated ITE in Step I. GUIDE is used to identify treatment effect patterns because of its reliable and robust performance in pattern recognition (Loh, 2002).

**Figure 1. A flow chart of the GUIDE pattern recognition algorithm in the setting of finding heterogeneous patterns of ITE**

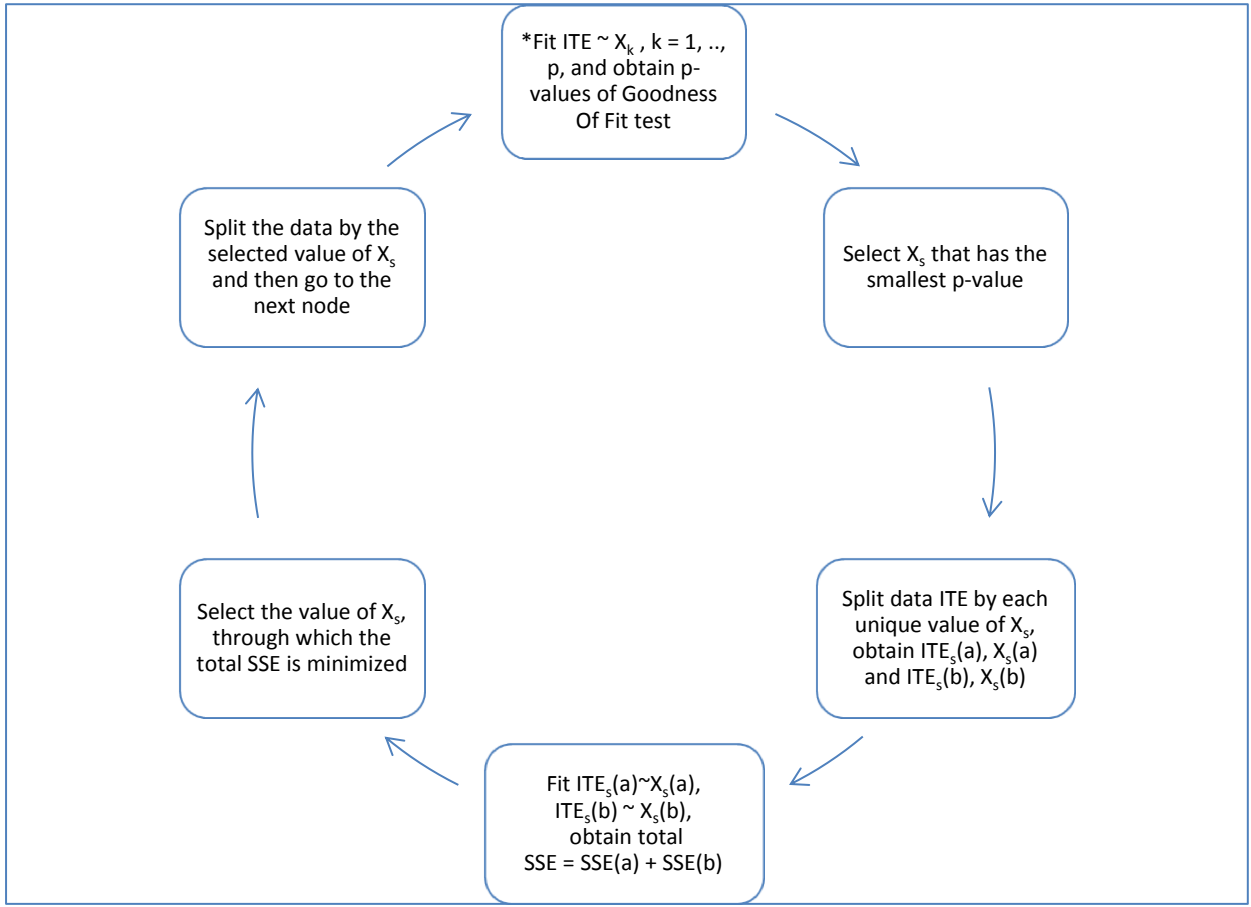

*: The starting point of the procedure

Figure 1 shows the procedures of GUIDE, in which ITE is the outcome, $X_k$ is the $k^{th}$ covariate, p is the total number of covariates. $X_s$ is the covariate that has the smallest p-value from fitting the univariate models ITE ~ $X_{kl}$ k = 1, …, p. $ITE_s(a)$, $ITE_s(b)$, $X_s(a)$ and $X_s(b)$ are the outcomes and covariate $X_s$ that were split by value $x_s$, respectively. SSE(a) and SSE(b) are the two SSEs obtained from two model fittings of $ITE_s(a)$, $ITE_s(b)$, $X_s(a)$ and $X_s(b)$ (Loh, 2002). Once GUIDE has identified the subgroups, the mean ITE difference for each subgroup is calculated. According to the algorithm, we can interpret the mean ITE difference as the difference between the outcome if the patient received treatment and the outcome if the patient received placebo, while adjusting for all the patient's covariates. When the original outcome variable Y is binary (0 or 1), the above algorithm is used to predict the probability of Y = 1, which will be ITE. And thus the interpretation of ITE difference would be in terms of the difference of two probabilities: probability of Y = 1 if the patient received treatment and probability of Y = 1 if the patient received placebo.

## Simulations

### Set-up

In order to evaluate the statistical properties of the VG method and to compare it to other methods, we performed a simulation study.

The targeted subgroup is the one with the corresponding signature(s) decided by the predictive variable interacting with the treatment variable. In other words, the treatment effect observed in the targeted subgroup is larger than that observed outside the targeted subgroup due to the interaction between the predictive variable and the treatment variable.

Thus, we first defined the true subgroup in our simulation. The signature was decided through predictive variable(s). To simplify, we considered cases where there was only one predictive variable, $X_{pred}$, in our simulation. Moreover, we defined one prognostic variable, $X_{prog}$, which had only main effect to the outcome but no interaction with the treatment.

We defined T as the treatment variable and Y as the outcome variable. In addition, we created $\mathbf{Z_{n,p}}$, a matrix containing p variables that are independent with the response and treatment variables.

In our simulation, we used the following models to generate the treatment vector and a covariate matrix:

$$T\ (0\ \text{or}\ 1) \sim \text{BIN}(n, 0.5)$$

$$\mathbf{X_{n,p+2}} = \begin{pmatrix} x_{\text{pred},1} & x_{\text{prog},1} & \\ \vdots & \vdots & \mathbf{Z_{n,p}} \\ x_{\text{pred},n} & x_{\text{prog},n} & \end{pmatrix} \sim \text{MVN}(0, \Sigma_{p+2,p+2})\ \text{if all}\ \mathbf{X_{n,p+2}}\ \text{are continuous and}$$

$$\Sigma_{p+2,p+2} = \begin{pmatrix} 1 & 0.5 & \cdots & 0.5 \\ 0.5 & 1 & \cdots & 0.5 \\ \vdots & \vdots & \ddots & \vdots \\ 0.5 & \cdots & 0.5 & 1 \end{pmatrix}$$

For the case where we simulated binary X for $X_{\text{pred}}$, $X_{\text{prog}}$ and/or $\mathbf{Z_{n,p}}$, we first simulated

$$p_x \sim \text{Beta}(2,3)$$

and we used

$$X \sim \text{BIN}(n, p_x)$$

to generate the values for binary covariates. And for outcome variable Y, we used the following models:

when Y is continuous:

$$Y = \beta_{\text{pred}} \times I(X_{\text{pred}} > x_0) \times T + \beta_{\text{prog}} \times X_{\text{prog}} + \beta_{\text{trt}} \times T + e$$

$$e \sim N(0, 0.25)$$

when Y is binary:

$$\mu = \beta_{pred} \times I(X_{pred} > x_0) \times T + \beta_{prog} \times X_{prog} + \beta_{trt} \times T$$

$$p_y = \frac{\exp(\mu)}{1 + \exp(\mu)}$$

$$Y \sim BIN(n, p_y)$$

Where $x_0$ is the cut-off point, in our simulation, we defined $x_0$ as the mean of $X_{pred}$ and $e$ is the noise that follows a normal distribution with mean 0 and variance 0.25. We'd like to simulate a dataset close to that from an observed clinical trial. Since we utilize a clinical trial with approximately 15 covariates, and 400 subjects, the resulting simulated datasets additionally contained 13 noise variables (p = 13), and 400 subjects (n = 400) within each of the iterations.

Therefore, the dataset contains all the components below:

$$Y = \begin{pmatrix} y_1 \\ \vdots \\ y_{400} \end{pmatrix} \quad T = \begin{pmatrix} t_1 \\ \vdots \\ t_{400} \end{pmatrix} \quad X = \begin{pmatrix} x_{pred,1} & x_{prog,1} & \\ \vdots & \vdots & Z_{n,p} \\ x_{pred,400} & x_{prog,400} & \end{pmatrix}$$

To simulate different scenarios, we selected different values for $\beta_{pred}$, $\beta_{prog}$ and $\beta_{trt}$. The different scenarios that we have simulated are summarized in Table 1.

Table 1. Simulation Scenarios

| Scenarios | $\beta_{pred}$ | $X_{pred}$ | $\beta_{prog}$ | $X_{prog}$ | $\beta_{trt}$ | Noise Variables (p = 13) |
|---|---|---|---|---|---|---|
| No Prognostic | 0.5 | Continuous | 0 | None | 0.2 | Continuous |
| No Prognostic Mix | 0.2 | Binary | 0 | None | 0.2 | Binary/ Continuous* |
| Prognostic | 0.5 | Continuous | 0.5 | Continuous | 0.2 | Continuous |

*Includes 1 binary variable and 12 continuous variables.

Through the three scenarios defined above, we used the following metrics to compare VG, Gi and VT methods:

1) Type I error: probability of identifying a subgroup when there are no subgroups.

2) Power: probability of identifying a subgroup when there is a subgroup.

3) Conditional true discovery rate: conditional probability of correctly identifying the predictive variable when a subgroup is identified.

For the purpose of fair comparison, we compared the power and true discovery rate for the three methods under the same type I error rate. We have simulated 500 iterations for each of the scenarios.

**Results**

According to the simulation results (Figure 2), all three methods behave similarly and demonstrate above 90% power and almost 100% conditional true discovery rate under most of Type I error rates when the predictive variable was continuous and there was no prognostic effect.

In the scenario where the predictive variable was binary with an effect size of 0.2 and one of the noise covariates was also binary, Gi demonstrated higher power and conditional true discovery rate than the other two methods, especially when the Type I error was controlled between 0 and 0.4.

When the prognostic effect was added to the simulation as a continuous variable, Gi had lower power and similar or lower conditional true discovery rate compared to the other two methods.

Since the VG and VT methods are very similar with respect to the background framework, these two methods behave very much alike. However, one can still notice about 5% improvement with VG compared to VT in the simulation results regarding the conditional true discovery rate, especially in the 'Prognostic' scenario.

**Figure 2. Plot of Power (Left) and Conditional True Discovery Rate (Right) vs Type I error**

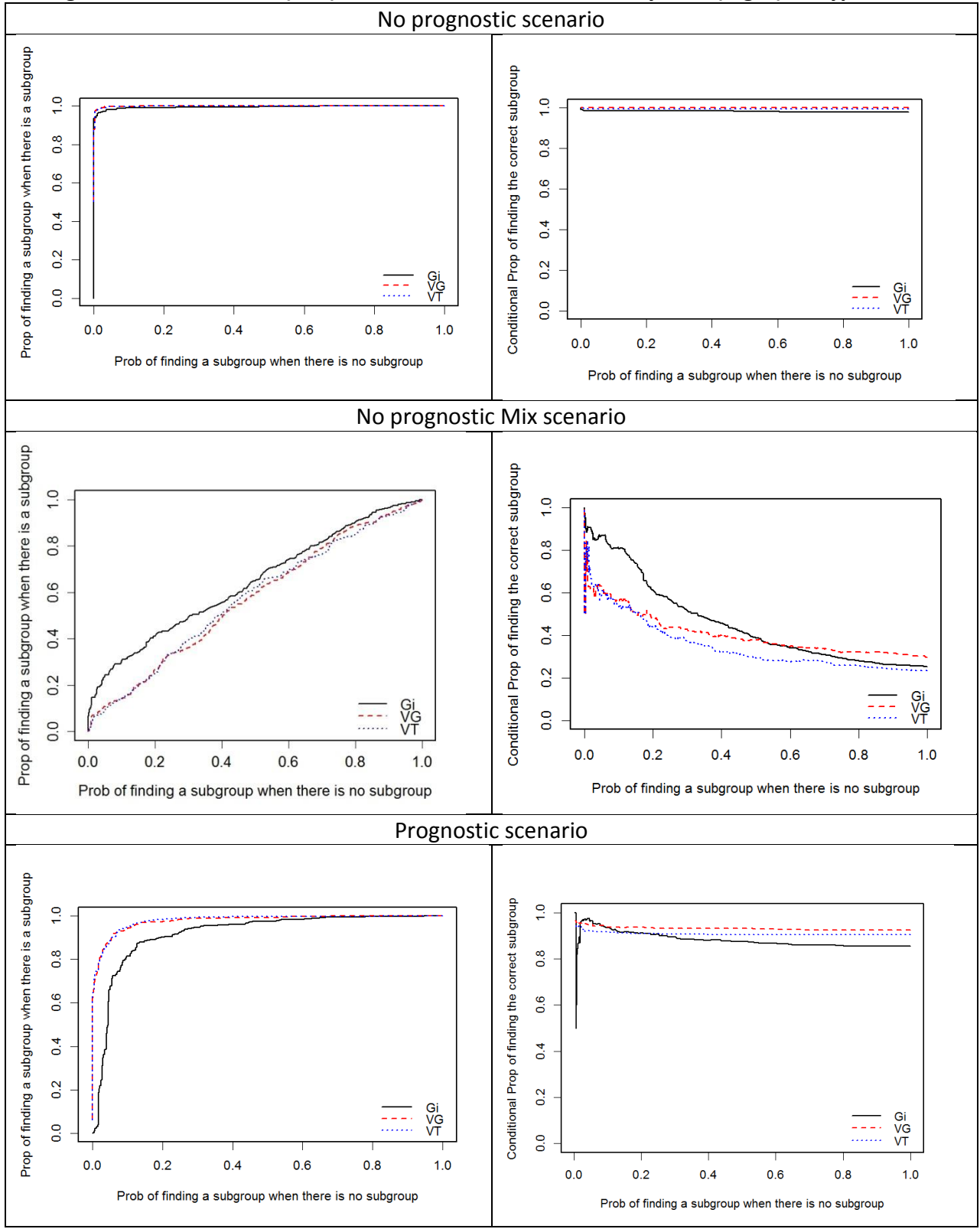

# Case study

**Type I error control**

In this paper, we define Type I error as the probability of identifying a subgroup when there are no subgroups. When conducting an analysis using real data, it can be challenging to obtain Type I error control as if the analysis was conducted using simulated data. Hence, we implemented the permutation method for the analysis involving real data.

Specifically, we break the association between treatment and covariate as well as between treatment and response; while keeping the association between covariate and response. In other words, we eliminated the predictive effect while keeping the prognostic effect. Other ways to control Type I error include the Šidák-based multiplicity adjustment method (Hochberg & Tamhane, 1987) as implemented in SIDES (Lipkovich, et al., 2011) and some more complicated permutation methods described in Foster et al (Foster, et al., 2016).

**Bootstrap**

In the analysis involving real data, one can easily obtain a naïve confidence interval for the point estimate of the treatment effect. However, such a confidence interval may not be valid because it does not take into consideration uncertainties from the selection of the predictive variable and the split value of the predictive variable. In the case study section of this paper, we implement a more complicated bootstrap method that is proposed by Loh et al (Loh, et al., 2015) to obtain confidence intervals for the point estimate of the treatment effect.

The bootstrap sample was drawn (with replacement) from the original dataset with the same size, and the VG method was applied. However, we ignored the identified signature based on the bootstrap sample. Instead, we obtained newly predicted ITE, which are different from ITE predicted during the first step of VG method. The new ITE can be obtained directly from GUIDE. Then, by using the identified signature of the subgroup from the original dataset, the bootstrap sample can be separated into subgroups and the mean of new ITE for these subgroups can be obtained. After these procedures have been repeated B times (B = 500 in our case), the distribution and the confidence interval of the mean of new ITE for the identified subgroup can be obtained.

**Application**

We applied the VG method to a real world example from a clinical study evaluating an experimental treatment for patients with Alzheimer's disease. The endpoint was the change from baseline to week 12 in ADAS-Cog 11 subscale score (0 to 70), which measures the change in severity of the disease. Thus, at the end of week 12, negative changes indicate improvement from baseline. There were two treatment arms: experimental treatment and placebo. In this case, the ITE for patient $i$ would be calculated as

$$\text{ITE}_i = y_i | Placebo, X_i - y'_i | Treatment, X_i$$

so that the larger the ITE, the better the treatment effect compared to placebo.

We have included 17 covariates after consulting with medical professionals, including but not limited to age, sex, race, baseline Mini-Mental State Examination (MMSE, a disease staging measure, range 0 – 30), the change of ADAS-Cog 11 subscale score from screening to baseline, and Apolipoprotein E4 (APOE4).

Since there were total of three datasets in this project, and we needed to control the Type I error while analyzing the data. Thus, we followed the steps below to conduct the subgroup identification analysis on the first two datasets.

1) Use permutation method on the first dataset to find the Type I error control;
2) Analyze the first dataset while controlling the Type I error, identify the signature(s); and
3) Find the subgroup in the second dataset according to the signature(s) identified in step 2, and evaluate the treatment effect in the subgroup to see if it differs from the other subgroup.

Unfortunately, when the Type I error was controlled at the 0.05 level, no subgroup was identified. Thus, we ignored Type I error control allowing for exploration of results that can be found. As shown in Figure 3, the covariate 'Years Since Onset of the Symptom' (YearOnset) was found as the predictive variable with a cut-off value at 3.55 years.

**Figure 3. Subgroup Identified From the First Dataset**

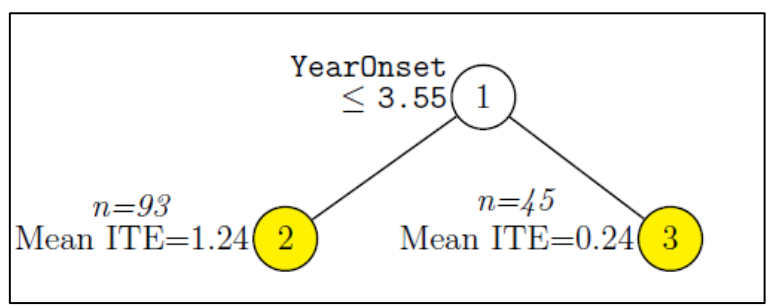

The cartoon in Figure 3 can be interpreted as follows (GUIDE reference): group '1' contains the overall patient population from this dataset. The patients who satisfy the criteria 'Years Since Onset of the Symptom ≤ 3.55 years' are classified into subgroup '2' (go along with the left line to circle 2), otherwise, the patients are classified into subgroup '3' (go along with the right line to circle 3).

According to the result, subgroup '2' had mean ITE = 1.24 with sample size 93, which means, on average, a larger treatment effect was observed in this subgroup of patients compared to the treatment effect observed in the rest of the patients (mean ITE = 0.24, n = 45). However, since Type I error was not controlled, we attempted to validate this result by observing the treatment effect in the subgroup of patients who were selected according to the identified signature ('Years Since Onset of the Symptom ≤ 3.55 years') from the second dataset, and compared it with the treatment effect observed from the first dataset.

Table 2. Estimated Treatment Effects

| Patient Group | Estimated Treatment Effect | |
|---|---|---|
| | First dataset | Second dataset |
| Overall | 0.91 | 0.07 |
| YearOnset ≤ 3.55 | 1.24 | 0.00 |
| YearOnset > 3.55 | 0.24 | 0.15 |

As shown in Table 2, the observed treatment effect in the subgroups from the first dataset cannot be replicated in the subgroups from the second dataset by using the same signature. Therefore, the result obtained from the VG method was not valid in this case given there was no Type I error control.

The third dataset in this project had a larger sample size than either of the first two datasets. Also, the experimental drug (Treatment) utilized in this dataset was different from the one utilized in the first two datasets. We applied the VG method on this dataset while controlling the Type I error at 0.05, and one covariate was identified as the predictive variable with a cut-off 20.

**Figure 4. Subgroup Identified From the Third Dataset**

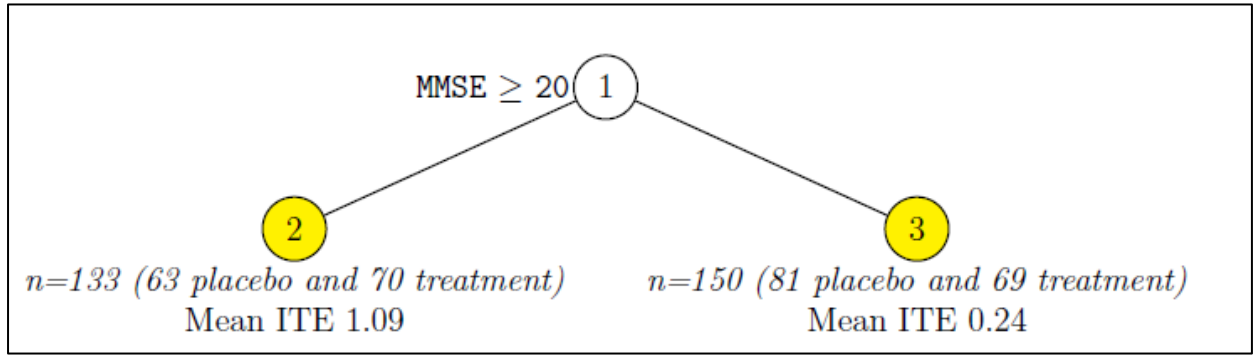

As shown in Figure 4, subgroup '2' was identified with a larger ITE compared to subgroup '3', although subgroup '3' had larger sample size. The identified covariate was MMSE, and the ITE appears to represent a clinically meaningful difference (Perneczky, et al., 2006). The cut-off value 20 is suggested to be used as the separation of disease staging between moderate (≤20) and mild (>20). For this dataset, we have also calculated the 95% confidence interval in order to obtain an estimation of the validity of the results.

**Table 3. Estimated Treatment Effects and 95% CI**

| Subgroup | Mean Effect | 95% CI |
|---|---|---|
| MMSE ≥ 20 (group '2') | 1.09 | (-0.11, 2.29) |
| MMSE < 20 (group '3') | 0.24 | (-1.19, 1.67) |
| Difference between Above Two Subgroups | 0.85 | (-0.79, 2.50) |

As shown in Table 3, although the treatment effect was numerically different from 0, the 95% confidence intervals for both subgroups contain 0. However, the 95% confidence interval in subgroup '2' is suggestive of a positive trend, while that in subgroup '3' is not. Moreover, the 95% confidence interval for the difference of the treatment effects between the two subgroups is also suggestive of a positive effect favoring subgroup 2 over subgroup 3. Therefore, these results were felt to be clinically meaningful. Additional exploration (i.e., studies) may be necessary to demonstrate whether this is truly a clinically meaningful effect.

## Discussion

Precision medicine attempts to improve the safety and/or efficacy of a drug by tailoring the treatment according to the patient's characteristics. Subgroup identification is a critical step to realize the potential of precision medicine. However, current realizations of subgroup analysis in clinical trials are often limited within pre-defined subgroups. The current state of conducting analyses according to pre-defined subgroups while ignoring Type I error control may result in true predictive variable(s) and/or true cut-off value(s) being missed. Some data-mining based subgroup identification methods also exist. Most of these methods are used to prospectively search for subgroups given a dataset. By using data mining techniques, one can avoid pitfalls of common one variable at a time subgroup analyses.

In this paper, we have proposed a novel method of prospective subgroup identification, the VG method, which combines the advantages of two existing methods (i.e., Virtual Twins and GUIDE). However, the VG method is not a simple combination of the two methods, it replace the CART part of the VT method by GUIDE. In other words, the VG method first calculates the Individual Treatment Effect (ITE) according to the counterfactual concept in causal inference; it then applies GUIDE to identify the subgroup(s) based on the ITE. Results from our simulation studies show that the VG method outperforms Virtual Twins when there are binary and continuous covariates in the data and also outperforms GI when prognostic effect is as strong as predictive effect. The key advantage of the VG method compared to Gi is that it targets directly on the treatment effect and can identify a predictive variable in the presence of a prognostic effect. Also, the VG method has less potential for selection bias when compared to Virtual Twins given the latter's reliance on CART. However, in our simulation, we have assumed there is only one predictive variable with one cut-off value due to the limitation of the tools we are using. In fact, there could be more than one predictive variable and there can be more than one cut-off value for a predictive variable in a given dataset.

Through the case study, without Type I error control, the identified subgroup is not valid and the results cannot be reproduced. When the Type I error is controlled, although the identified subgroup is not statistically significant based on the 95% confidence intervals calculated using the bootstrap method, it demonstrates a trend related to the treatment effect that might be clinically meaningful. In other words, with conservatively controlled Type I error, the results might not be statistically significant, but might provide some clinically helpful information. In this case, additional research (i.e., clinical trials) would be needed to confirm this result.

Our work has provided a clearly defined framework to compare three different subgroup identification methods according to type I error control, power and the conditional true discovery rate. It also provides two applications for controlling type I error and estimating 95% confidence intervals in the analysis of a real dataset, which are permutation and bootstrap methods, respectively. The performance of VG method relies on datasets, case by case simulation is suggested to be tailored to the study. Generalization the conclusion to other studies should be careful.

Our future work involves the improvement of the prediction accuracy when calculating the ITE, which is a critical factor that impacts the performance of the VG method. Moreover, we are trying to extend the VG method to both binary and time-to-event endpoints.